\documentclass[prd,superscriptaddress,twocolumn,nofootinbib,showpacs]{revtex4-1}
\usepackage{amsfonts,amssymb,amsmath}
\usepackage{mathrsfs}
\usepackage{color}
\usepackage{graphicx}
\usepackage{dcolumn}
\usepackage{bm}

\usepackage{pslatex}
\usepackage[T1]{fontenc}
\usepackage[latin1]{inputenc}
\usepackage{graphicx}
\usepackage{epsfig}
\usepackage{longtable}
\usepackage{float}
\usepackage{calc}
\usepackage{ifthen}

{
{
{
\newcommand{\bea}{\begin{eqnarray}}
\newcommand{\eea}{\end{eqnarray}}

\newcommand{\nc}{\newcommand}
\nc{\renc}{\renewcommand}
\nc{\eqs}[2]{\mbox{Eqs.~(\ref{#1},\,\ref{#2})}}
\nc{\eq}[1]{\mbox{Eq.~(\ref{#1})}}
\nc{\figs}[2]{\mbox{Figs.~(\ref{#1},\,\ref{#2})}}
\nc{\fig}[1]{\mbox{Fig~.(\ref{#1})}}
\nc{\be}[1]{\begin{equation} \mbox{$\label{#1}$}}
\nc{\ee}{\vspace{0.1cm}\end{equation}}

\newcommand{\bean}{\begin{eqnarray*}}
\newcommand{\eean}{\end{eqnarray*}}

\def\bfx{{\bf x}}
\def\bfk{{\bf k}}

\begin{document}
\title{Chaotic initial conditions for nonminimally coupled inflation via a conformal factor with a zero}
\author{Jinsu Kim}
\email{kimjinsu@kias.re.kr}
\affiliation{Quantum Universe Center, Korea Institute for Advanced Study, Seoul 02455, Korea}
\author{John McDonald}
\email{j.mcdonald@lancaster.ac.uk}
\affiliation{Department of Physics, Lancaster University, Lancaster LA1 4YB, United Kingdom}

%%%%%%%%%%%%%%%%%%%%%%%%%%%%%%%%%%%%%%%%%%%%%%%%%
\begin{abstract}

Nonminimally coupled inflation models based on a nonminimal coupling $\xi \phi^{2} R$ and a $\phi^{4}$ potential are in excellent agreement with the scalar spectral index observed by Planck. Here we consider the modification of these models by a conformal factor with a zero. This enables a nonminimally coupled model to have a Planck-scale potential energy density at large values of the inflaton field, which can account for the smooth, potential-dominated volume that is necessary for inflation to start. We show that models with a conformal factor zero generally predict a correlated increase of the spectral index $n_{s}$ and tensor-to-scalar ratio $r$. For values of $n_{s}$ that are within the present $2-\sigma$ bounds from Planck, modification by $\Delta r$ as large as 0.0013 is possible, which is large enough to be measured by next generation cosmic microwave background polarization satellites.  

\end{abstract}
%%%%%%%%%%%%%%%%%%%%%%%%%%%%%%%%%%%%%%%%%%%%%%%%%
 
\maketitle

%%%%%%%%%%%%%%%%%%%%%%%%%%%%%%%%%%%%%%%%%%%%%%%%%
\section{Introduction}
\label{sec:intro}
%%%%%%%%%%%%%%%%%%%%%%%%%%%%%%%%%%%%%%%%%%%%%%%%%

Nonminimally coupled scalar field inflation models of the type first proposed by Salopek \textit{et al.} \cite{Salopek:1988qh} have the great advantage of being able to use $\phi^{4}$ scalar potentials with a self-coupling $\lambda$ of magnitude typical of particle physics models. Examples include Higgs inflation \cite{Bezrukov:2007ep}, inflation models based on dark matter gauge singlet scalars \cite{Lerner:2009xg, Kahlhoefer:2015jma, Kim:2014eia}, and supersymmetric extensions of Higgs inflation \cite{Kawai:2014gqa, Kawai:2015ryj}. Nonminimally coupled inflation predicts $n_{s} \approx 1 -2/N - 3/N^2  \approx 0.966$  and $r \approx  12/N^2  \approx 3.3\times 10^{-3}$ at $N = 60$ (where $N$ is the number of $e$-foldings in the Einstein frame), in very good agreement with the observed spectral index, $n_{s} = 0.9688 \pm 0.0061$ (68$\%$ C.L., Planck TT + lowP + lensing), and easily consistent with the upper bound on the tensor-to-scalar ratio,  $r_{0.002} < 0.114$ (95$\%$ C.L., Planck TT + lowP + lensing)~\cite{Ade:2015lrj}.

Inflation requires a smooth potential-dominated initial state over a horizon volume. Therefore, in order to be a complete theory, inflation requires a physical mechanism that can explain this initial state. A favored approach to creating the initial conditions for inflation is to assume that the Universe started in a chaotic initial state with Planck-scale energy density $\mathcal{O}(M_{{\rm P}}^{4})$ \cite{Linde:1983gd, Linde:1984ir}. The initial classical state is expected to be generated from a quantum fluctuation which has Planck energy density and size around the scale of the horizon $H^{-1} \approx M_{{\rm P}}^{-1}$. The initial classical energy density of the Planck scale fluctuation is assumed to be distributed roughly equally between the kinetic, gradient and potential energy densities of the scalar field. The potential energy density can then quickly come to dominate as the Universe subsequently expands, thereby creating the required smooth potential-dominated initial state on the scale of the horizon. 

For this mechanism to work, the potential energy density must be able to reach the Planck energy density for some value of the scalar field $\phi$.  Therefore plateau inflation models with $V(\phi) \ll M_{{\rm P}}^{4}$ cannot become potential-dominated during an initial Planck density era. There are a number of ways that the smooth potential-dominated initial state on scale of the horizon can be produced for a plateau potential. One way, which is the focus of our discussion here, is to modify the potential such that it increases as the inflaton field $\phi$ increases and reaches the Planck energy density. Another proposal is for a smooth patch to be produced during the chaotic era which has the form of an open Universe, with a negative curvature term which dominates the Friedmann equation \cite{Guth:2013sya}. In this case the Hubble radius during the subsequent expansion satisfies $H^{-1} \propto a$, where $a$ is the scale factor, and so a smooth horizon-sized patch at the Planck density will expand to a smooth horizon-sized patch at the onset of plateau inflation. A different approach, which does not rely on a chaotic initial state, is to have a contracting era which precedes the expanding era. This can be achieved by a generalization of the nonminimal coupling and potential \cite{McDonald:2015cqe}\footnote{For more a recent proposal to address the initial condition problem of plateau inflation, see Ref.~\cite{Dimopoulos:2016yep}.}.

In this paper we will focus on the idea that the potential will increase to the Planck energy at large $\phi$. The usual way to achieve this is to add nonrenormalizable higher-order terms to the potential. In the case of nonminimally coupled inflation models, there is an alternative approach, which we will explore here. This is to consider a conformal factor with a zero. In this case, the Einstein-frame potential will have a pole and so will rapidly increase to the Planck energy density as $\phi$ approaches the pole. This class of models may be regarded as a minimal modification of the standard nonminimally coupled inflation model, in the sense that they modify only the nonminimal coupling of the scalar particle to gravity and leave the particle physics model unchanged.

The paper is organized as follows. In Sec. \ref{sec:NMinf} we review nonminimally coupled inflation models and the results for the standard nonminimally coupled model. In Sec. \ref{sec:NMpole} we introduce a class of models which have a conformal factor with a zero at large $\phi$ and which reduce to the standard nonminimal model at small $\phi$. In Sec. \ref{sec:NMpoleICs} we discuss the evolution of these models from chaotic initial conditions. In Sec. \ref{sec:NMpoleCOs} we discuss the possibility of detecting a deviation of $n_{s}$ and $r$ from their standard nonminimally coupled inflation model values. In Sec. \ref{sec:con} we present our conclusions.

%%%%%%%%%%%%%%%%%%%%%%%%%%%%%%%%%%%%%%%%%%%%%%%%
\section{Nonminimally Coupled Inflation Models}
\label{sec:NMinf}
%%%%%%%%%%%%%%%%%%%%%%%%%%%%%%%%%%%%%%%%%%%%%%%%

In general, a nonminimally coupled scalar model is described by an action with nonminimal coupling $F(\phi)$, a generic kinetic coupling $Z(\phi)$, and a generic potential $V_{{\rm J}}(\phi)$,
\begin{align}
	S_{{\rm J}} = 
	\int d^{4}x \sqrt{-g_{{\rm J}}} \left[
		\frac{M_{{\rm P}}^{2}}{2}\Omega^{2}(\phi)R_{{\rm J}}
		-\frac{1}{2}g_{{\rm J}}^{\mu\nu}Z(\phi)
		\partial_{\mu}\phi\partial_{\nu}\phi
		-V_{{\rm J}}(\phi)
	\right]\,,
\end{align}
where $\Omega^{2}=1+F(\phi)$ is the conformal factor. The subscript J stands for the Jordan frame.
The action in the Einstein frame, denoted by the subscript E, can be obtained via the Weyl rescaling,
\begin{align}
	g_{{\rm J}}^{\mu\nu} \rightarrow
	g_{{\rm E}}^{\mu\nu}
	= \Omega^{-2}g_{{\rm J}}^{\mu\nu}\,,
\end{align}
which gives
\begin{align}
	S_{{\rm E}} =
	\int d^{4}x \, \sqrt{-g_{{\rm E}}} \, \left[
		\frac{M_{{\rm P}}^{2}}{2}R_{{\rm E}}
		-\frac{1}{2}g_{{\rm E}}^{\mu\nu}
		\partial_{\mu}\varphi
		\partial_{\nu}\varphi
		-V_{{\rm E}}(\varphi)
	\right]\,,
\end{align}
where $\varphi$ is the canonically normalized field and $V_{{\rm E}}$ is the Einstein-frame potential, which are respectively related to the field $\phi$ and the Jordan-frame potential $V_{{\rm J}}$ as follows:
\begin{align}
	\left(
		\frac{d\varphi}{d\phi}
	\right)^{2}
	&=
	\frac{Z}{\Omega^{2}}
	+\frac{3M_{{\rm P}}^{2}}{2\Omega^{4}}
	\left(
		\frac{d\Omega^{2}}{d\phi}
	\right)^{2}
	\,,\label{eqn:JErelfield}
	\\
	V_{{\rm E}}
	&=
	\frac{V_{{\rm J}}}{\Omega^{4}}
	=
	\frac{V_{{\rm J}}}{(1+F)^{2}}
	\,.\label{eqn:JErelpot}
\end{align}
In the following we will mostly state results in terms of the Jordan-frame field $\phi$, even though inflation is analyzed in the Einstein frame in terms of the canonically normalized field $\varphi$. 

In terms of the Einstein-frame potential $V_{{\rm E}}$ and the canonically normalized field $\varphi$, the cosmological observables are expressed in terms of the potential slow-roll parameters $\epsilon$ and $\eta$, defined by 
\begin{align}
	\epsilon =
	\frac{M_{{\rm P}}^{2}}{2}\left(
		\frac{dV_{{\rm E}}/d\varphi}{V_{{\rm E}}}
	\right)^{2}
	\,,\qquad
	\eta =
	M_{{\rm P}}^{2}\left(
	\frac{d^{2}V_{{\rm E}}/d\varphi^{2}}{V_{{\rm E}}}
	\right)
	\,.\label{eqn:SRparams}
\end{align}
In the following we will evaluate the cosmological observables at $N=60$, where $N$ is the number of $e$-folds in the Einstein frame:
\begin{align}
	N &= \int_{t_{{\rm i}}}^{t_{{\rm e}}} \, H \, dt
	\approx
	-\frac{1}{M_{{\rm P}}^{2}}
	\int_{\varphi_{{\rm i}}}^{\varphi_{{\rm e}}} \, d\varphi
	\,
	\frac{V_{{\rm E}}}{dV_{{\rm E}}/d\varphi}
	\nonumber\\
	&=
	\frac{1}{M_{{\rm P}}}
	\int_{\phi_{{\rm e}}}^{\phi_{{\rm i}}}
	\, d\phi \,
	\frac{1}{\sqrt{2\epsilon}}
	\frac{d\varphi}{d\phi}
	\,,
\end{align}
where $H = \dot{a}/a$ is the expansion rate defined in the Einstein frame, where a flat Friedmann-Robertson-Walker metric in the Einstein frame is assumed with scale factor $a$ and time coordinate $t$.

We next review the relevant results of what has come to be the standard nonminimally coupled scalar inflation model, which we will refer to as the ``standard nonminimal model'' for convenience.
In these models the nonminimal coupling, the kinetic coupling, and the potential in the Jordan frame are specified by
\begin{align}
    \Omega^2  &= 
		1 + \xi_{2} \frac{\phi^{2}}{M_{{\rm P}}^{2}}
	\,,\qquad
	Z = 1\,, \qquad
	V_{{\rm J}} = \frac{\lambda}{4}\phi^{4}\,.
\end{align}
The Einstein-frame potential $V_{{\rm E}}$ and the relation between the canonically normalized field $\varphi$ and the original field $\phi$ are, according to Eqs.~\eqref{eqn:JErelfield} and \eqref{eqn:JErelpot}, given by
\begin{align}
	V_{{\rm E}} &= \frac{\lambda\phi^{4}}{4(1+\xi_{2}\phi^{2}/M_{{\rm P}}^{2})^{2}}
	\,,\\
	\frac{d\varphi}{d\phi}
	&=
	\frac{\sqrt{
		1+(1+6\xi_{2})\xi_{2}\phi^{2}/M_{{\rm P}}^{2}
	}}{1+\xi_{2}\phi^{2}/M_{{\rm P}}^{2}}
	\,.
\end{align}
The slow-roll parameters \eqref{eqn:SRparams} are 
\begin{align}
	\epsilon
	\approx
	\frac{4}{3\xi_{2}^{2}}\left(
		\frac{M_{{\rm P}}}{\phi}
	\right)^{4}
	\,,\qquad
	\eta
	\approx
	-\frac{4}{3\xi_{2}}\left(
		\frac{M_{{\rm P}}}{\phi}
	\right)^{2}
	\,,\label{eqn:SRparamsSNMinf}
\end{align}
where we have taken the large-field limit, $\phi \gg M_{{\rm P}}/\sqrt{\xi_{2}}$.
The end of inflation is then specified by the condition $\epsilon \approx 1$, which gives
\begin{align}
	\phi_{{\rm e}}
	\approx
	\left( \frac{4}{3} \right)^{1/4}
	\frac{M_{{\rm P}}}{\sqrt{\xi_{2}}}
	\,.
\end{align}
In the large-field limit, the number of $e$-folds is given by
\begin{align}
	N(\phi) \approx
	\frac{3\xi_{2}}{4}\left(
		\frac{\phi^{2}}{M_{{\rm P}}^{2}}
		-\frac{\phi_{{\rm e}}^{2}}{M_{{\rm P}}^{2}}
	\right)\,.
\end{align}
Therefore, at $\phi^2 \gg \phi_{e}^2$, we have $\phi \approx \sqrt{4 N/3 \xi_{2}} M_{P}$. Thus $\phi(N = 60) \approx 9 M_{{\rm P}}/\sqrt{\xi_{2}}$. 

The slow-roll parameters \eqref{eqn:SRparamsSNMinf} at $N = 60$ take the values $\eta_{60} \approx -0.0165$ and $\epsilon_{60} \approx 0.000203$. The cosmological observables are then given by
\begin{align}
	\mathcal{P}_{s}
      \approx \frac{V_{{\rm E}}}{24\pi^{2}M_{{\rm P}}^{4}\epsilon}
	\approx
	5.19 \lambda/\xi_{2}^{2}
	\,, 
\end{align}
\begin{align}
     n_{s}
     \approx 1+2\eta - 6\epsilon
	\approx
	0.9659
	\,,\qquad
	r \approx 16\epsilon
	\approx
	0.003
	\,.
	\label{eqn:NMstandardCOs}
\end{align}
Using the Planck result \cite{Ade:2015lrj}, $\mathcal{P}_{s} \approx 2.2 \times 10^{-9}$, we then obtain the relation between the nonminimal coupling $\xi_{2}$ and the quartic coupling $\lambda$, $\xi_{2} = 4.88 \times 10^{4} \sqrt{\lambda}$.

%%%%%%%%%%%%%%%%%%%%%%%%%%%%%%%%%%%%%%%%%%%%%%%%
\section{Nonminimally Coupled Inflation with a Conformal Factor Zero}
\label{sec:NMpole}
%%%%%%%%%%%%%%%%%%%%%%%%%%%%%%%%%%%%%%%%%%%%%%%%

In this section we introduce a class of models which have a conformal factor with a zero. These models, to a good approximation, reduce to the standard nonminimal model at $\phi(N = 60)$, so preserving the successful prediction for $n_{s}$, while having a conformal factor with a zero at large $\phi$. In keeping with the absence of a linear term in the conformal factor of the standard nonminimal model, we will assume that there is a symmetry preventing terms which are odd in the $\phi$ field. 

We will therefore consider a general class of models where the nonminimal coupling of the standard model has a correction factor which is a function of $\phi^{2}/M_{{\rm P}}^{2}$,
\begin{align}
	\frac{\xi_{2}\phi^{2}}{M_{{\rm P}}^{2}}
	\rightarrow
	\frac{\xi_{2}\phi^{2}}{M_{{\rm P}}^{2}}
	\times
	f\left(
		\frac{\phi^{2}}{M_{{\rm P}}^2}
	\right)\,.
\end{align}
The function $f(\phi^{2}/M_{{\rm P}}^{2})$ must tend to 1 at small $\phi$ and is assumed to make the conformal factor equal zero at large $\phi$. Then $f(\phi^{2}/M_{{\rm P}}^{2})$ can be Taylor expanded at small $\phi^{2}/M_{{\rm P}}^{2}$ as 
\begin{align}
	f\left(
		\frac{\phi^{2}}{M_{{\rm P}}^{2}}
	\right)
	= 
	1 + a_{1} \frac{\phi^{2}}{M_{{\rm P}}^{2}}
	+ a_{2} \frac{\phi^{4}}{M_{{\rm P}}^{4}}
	+ \cdots
	\,,
\end{align}
where $a_{i}$ are expected to be of order 1. Thus at $\phi^{2} \ll M_{{\rm P}}^{2}$ we have
\begin{align}
	\Omega^{2} 
	= 
	1
	+\frac{\xi_{2} \phi^{2}}{M_{{\rm P}}^{2}}
	\times f\left(
		\frac{\phi^{2}}{M_{{\rm P}}^{2}}
	\right)
	= 
	1
	+\frac{\xi_{2} \phi^{2}}{M_{{\rm P}}^{2}}
	+\frac{a_{1} \xi_{2} \phi^{4}}{M_{{\rm P}}^{4}}   
	+\cdots
	\,.
	\label{eqn:generalcase}
\end{align}
In general $a_{1}$ could be either positive or negative. However, since this term will dominate the initial deviation from the plateau potential as $\phi$ increases, $a_{1}$ must be negative, in order to prevent the potential in the Einstein frame from developing a local minimum, where $\phi$ would become trapped as it rolls in from the chaotic initial state.
Therefore, at small $\phi/M_{{\rm P}}$, the leading order contributions to $\Omega^{2}$ will generally have the form 
\begin{align}
	\Omega^2 = 
		1 + \xi_{2} \frac{\phi^{2}}{M_{{\rm P}}^{2}}
		-\xi_{4}\frac{\phi^{4}}{M_{{\rm P}}^{4}} + ...
	\,.
\end{align}
where we expect $\xi_{4}  = |a_{1}| \xi_{2} = \mathcal{O}(1) \times  \xi_{2}$. Thus $\xi_{4} \sim \xi_{2}$ is natural in these models. This will be important later for observable deviations from the predictions of the standard nonminimal inflation model. The observable predictions of this class of model depend only on the $\phi^{4}$ term in the expansion at small $\phi^{2}/M_{{\rm P}}^2$; therefore, they are independent of the precise form of $f(\phi^{2}/M_{{\rm P}}^{2})$.

As a specific example, we will consider a minimal model with a conformal factor zero. This is defined by
\begin{align}
	\Omega^{2} = 
		1 + \xi_{2} \frac{\phi^{2}}{M_{{\rm P}}^{2}}
		-\xi_{4}\frac{\phi^{4}}{M_{{\rm P}}^{4}}
	\,, \qquad
	Z = 1\,, \qquad
	V_{{\rm J}} = \frac{\lambda}{4}\phi^{4}\,.
\end{align}
This model is characterized by a single additional parameter, $\xi_{4}$. The conformal factor has a zero at a critical field value, $\phi_{{\rm c}}$, given by
\begin{align}
	\phi_{{\rm c}} = \frac{M_{{\rm P}}}{\sqrt{2\xi_{4}}}\left(
		\xi_{2} + \sqrt{\xi_{2}^{2}+4\xi_{4}}
	\right)^{1/2}
	\,.\label{eqn:phicrit}
\end{align}
For this model, the Einstein-frame potential $V_{{\rm E}}$ takes the form
\begin{align}
	V_{{\rm E}} &=
	\frac{\lambda \phi^{4}}
	{4(1+\xi_{2}\phi^{2}/M_{{\rm P}}^{2}
	-\xi_{4}\phi^{4}/M_{{\rm P}}^{4})^{2}}
	\,.\label{eqn:EFpotNMpole}
\end{align}
Thus the conformal factor results in a pole in the Einstein-frame potential at $\phi=\phi_{{\rm c}}$ \eqref{eqn:phicrit}. 
Therefore the potential energy density in the Einstein frame will approach the Planck energy density as $\phi$ increases, which allows chaotic initial conditions to be consistent with the model. (We expect the chaotic initial conditions to be determined in the Einstein frame, where the model is minimally coupled to gravity and where quantum fluctuations will become large when $H \sim M_{{\rm P}}$, where $H$ is the expansion rate in terms of the Einstein-frame scale factor.) The initial value of the field, $\phi_{{\rm IC}}$, is defined by $V_{{\rm E}}(\phi_{{\rm IC}}) = M_{{\rm P}}^{4}$, which implies that
\begin{align}
	\phi_{{\rm IC}} =
	\frac{M_{{\rm P}}}{\sqrt{2\xi_{4}}}\left[
		\xi_{2} - \frac{\sqrt{\lambda}}{2}
		+\sqrt{\left(
			\xi_{2} - \frac{\sqrt{\lambda}}{2}
		\right)^{2} + 4\xi_{4}}
	\right]^{1/2} \,. \label{eqn:phiChaoticIC}
\end{align}
Note that $\phi_{{\rm IC}} \approx M_{{\rm P}}$ when $\xi_{2}\sim \xi_{4}$ and $\xi_{2} \gg \sqrt{\lambda}$.

The canonically normalized field in the Einstein frame $\varphi$ is related to the field $\phi$ in this model via Eq.~\eqref{eqn:JErelfield},
\begin{align}
	\frac{d\varphi}{d\phi} &=
	\frac{
	1
	}{
	1+\xi_{2}\frac{\phi^{2}}{M_{{\rm P}}^{2}}
	-\xi_{4}\frac{\phi^{4}}{M_{{\rm P}}^{4}}
	} \times
	\nonumber\\
	&
	\left[
	1+(1+6\xi_{2})\xi_{2}\frac{\phi^{2}}{M_{{\rm P}}^{2}}
	-(1+24\xi_{2})\xi_{4}\frac{\phi^{4}}{M_{{\rm P}}^{4}}
	+24\xi_{4}^{2}\frac{\phi^{6}}{M_{{\rm P}}^{6}}
	\right]^{1/2}
	\,.
	\label{eqn:JErelfield2}
\end{align}
%%

%%%%%%%%%%%%%%%%%%%%%%%%%%%%%%%%%%%%%%%%%%%%%%%%
\section{Initial conditions in the minimal conformal factor zero model}
\label{sec:NMpoleICs}
%%%%%%%%%%%%%%%%%%%%%%%%%%%%%%%%%%%%%%%%%%%%%%%%

The expected chaotic initial condition for the subsequent classical evolution is $\varphi \sim \varphi_{{\rm IC}}$, with the canonically normalized field in the Einstein frame satisfying $(\dot{\varphi}_{{\rm IC}})^{2} \sim (\nabla{\varphi}_{{\rm IC}})^{2} \sim V_{{\rm E}}(\varphi_{{\rm IC}}) \sim M_{{\rm P}}^{4}$. 

The potential energy density approaches the Planck energy density when $\varphi \approx \varphi_{{\rm IC}}$ \eqref{eqn:phiChaoticIC}, at which point the potential in the Einstein frame is steep. It is therefore important to check that when the potential subsequently satisfies the slow-roll conditions, the kinetic and gradient energy densities from the initial evolution from the Planck density are not strongly dominant at this time. If the kinetic energy were strongly dominant at this time, then the field would enter into oscillations, with the possibility that perturbations could grow and come to dominate the potential, losing the smooth potential-dominated horizon volume which is necessary for plateau inflation to begin. Similarly, if the gradient energy were dominant at this time, then we would not have the smooth potential-dominated state on the scale of the horizon which is necessary for inflation to begin. 

We first check that the kinetic energy of the rolling field does not strongly dominate the potential energy when the slow-roll conditions are satisfied by the potential. Denoting the kinetic energy density of the field $\varphi$ by $\rho_{{\rm kin}} \equiv \dot{\varphi}^{2}/2$, where for now we neglect any inhomogeneities, the time derivative of the total energy density $\rho=\rho_{{\rm kin}}+V_{{\rm E}}$ is given by
\begin{align}
	\frac{d\rho}{dt}
	=
	\left[
		\ddot{\varphi}
		+\frac{dV_{{\rm E}}}{d\varphi}
	\right]\dot{\varphi}
	\,.
\end{align}
Using the inflaton field equation
\begin{align} 
\ddot{\varphi} + 3 H \dot{\varphi} = - \frac{dV_{{\rm E}}}{d\varphi} ~\, ,
\end{align}
one can easily show that
\begin{align}
	M_{{\rm P}}
	\frac{d\rho_{{\rm kin}}}{d\varphi}
	=
	\sqrt{3\rho}
	|\dot{\varphi}|
	-\sqrt{2\epsilon}V_{{\rm E}}
	\,.
	\label{eqn:deltakin1}
\end{align}
In this we have used $\dot{\varphi} = - |\dot{\varphi}|$, as $\varphi$ is rolling in towards the origin. Let us now assume that the kinetic energy density comes to dominate, so that $\rho \approx \rho_{{\rm kin}}$. The largest possible value of $\rho_{{\rm kin}}$ at any $\varphi$ will be obtained when the right-hand side of the equation equals zero. Therefore the maximum kinetic energy is
\begin{align} 
\rho_{{\rm kin\;max}} = \sqrt{\frac{\epsilon}{3}} V_{{\rm E}} \,.
\end{align}
The potential satisfies the slow-roll condition once $\epsilon \approx 1$, therefore at this time $\rho_{{\rm kin\;max}} \approx V_{{\rm E}}/\sqrt{3}$. This contradicts the assumption that the energy is dominated by the kinetic energy. Thus when the slow-roll conditions are satisfied we must have $\rho_{{\rm kin}} \lesssim V_{{\rm E}}$. The field will therefore rapidly enter potential-dominated slow-roll inflation once the slow-roll conditions on the scalar potential are satisfied. 

\begin{figure}[t]
\centering
\includegraphics[width=0.45\textwidth]{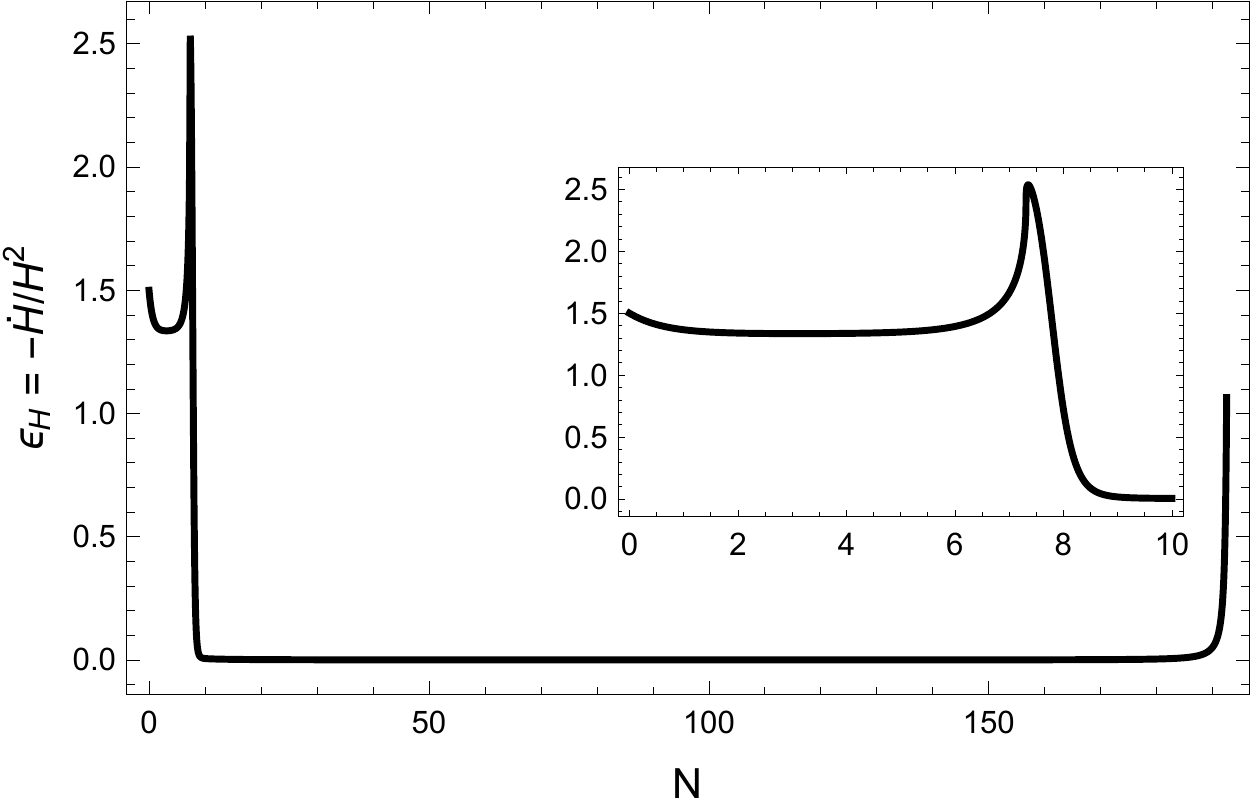}
\caption{\label{fig:epsH}
The Hubble slow-roll parameter $\epsilon_{H} = -\dot{H}/H^{2}$ as a function of the number of $e$-folds $N$.
After a short period of fast roll, the Hubble slow-roll parameter becomes smaller than unity at $N \approx 8$, resulting in slow-roll inflation. It later starts to increase near the end of inflation.
A sudden increase around $N \approx 7.5$ is due to the relation between $\varphi$ and $\phi$; see Eq.~\eqref{eqn:JErelfield2}.
In the analysis, we take $\xi_{4} \approx \xi_{2} \approx 3 \times 10^{4}$ and $\lambda = 0.5$. The generic behavior, however, is unaltered by different choices of parameters.
}
\end{figure}

We have also checked this conclusion numerically. Figure~\ref{fig:epsH} shows the Hubble slow-roll parameter, $\epsilon_{H} = -\dot{H}/H^{2}$, in terms of the number of $e$-folds\footnote{In the limit where the energy density is dominated by the potential, the potential slow-roll parameters $\eta$ and $\epsilon$ are related to the Hubble slow-roll parameters by $\eta = 2 \epsilon_{H} - \eta_{H}/2$ and $\epsilon = \epsilon_{H}$, where $\eta_{H} = \dot{\epsilon}_{H}/(H \epsilon_{H})$. Therefore both $\epsilon_{H}$ and $|\eta|_{H}$ must be less than 1 in order to satisfy the potential slow-roll condition $\{ \epsilon, |\eta| \} \lesssim 1$. We find that $\epsilon_{H}$ is somewhat larger than $|\eta|_{H}$; therefore, $\epsilon_{H} \lesssim 1$ defines the onset of slow-roll inflation in this model.} $N$.  After a short period of fast rolling, the Hubble slow-roll parameter becomes smaller than unity at $N \approx 8$, resulting in slow-roll inflation, and finally increases to unity at the end of inflation. The sharp increase in $\epsilon_{H}$ around $N \approx 7.5$ is due to the relation between $\varphi$ and $\phi$ \eqref{eqn:JErelfield2}.

We next consider the spatial fluctuations and gradient energy during the initial fast-roll period following the chaotic initial state. To model these fluctuations, we will consider the following field:
\begin{align}
 \varphi(\bfx, t) =  \overline{\varphi}(t) + \delta  \varphi (\bfx, t) = \overline{\varphi}(t) + \delta  \varphi_{k}(t) e^{i\bfk.\bfx}   \,.
\end{align}
The initial value of the homogeneous field $\overline{\varphi}(0)$ is assumed to be equal to $\varphi_{{\rm IC}}$, defined by $V_{{\rm E}}(\varphi) = M_{{\rm P}}^{4}$. (We set $t = 0$ and $a = 1$ initially.)
We have also assumed that there is a spatial fluctuation $\delta \varphi$ of comoving wavelength $\lambda = 2 H(0)^{-1}$, corresponding to $k = \pi H(0)$, and gradient energy density $M_{{\rm P}}^{4}$, which models the initial horizon-sized chaotic fluctuations. The initial values are then $\overline{\varphi}(0) \approx 23 M_{{\rm P}}$ [corresponding to $\phi_{{\rm IC}} \approx M_{{\rm P}}$ \eqref{eqn:phiChaoticIC}] and $\delta \varphi_{k}(0) \approx 0.55 M_{{\rm P}}$ (corresponding to $\rho_{{\rm grad}} \approx M_{{\rm P}}^{4}$). We subsequently evolve this as a classical mode, with $H$ determined by the total energy density inside the classical volume (which is assumed to be spherical), in order to model the expected inhomogeneity from the chaotic initial conditions. We also treat $\delta \varphi$ as a perturbation of $\overline{\varphi}$, which we find is consistent throughout, and therefore consider the background metric to be homogeneous.

\begin{figure}[t]
\centering
\includegraphics[width=0.45\textwidth]{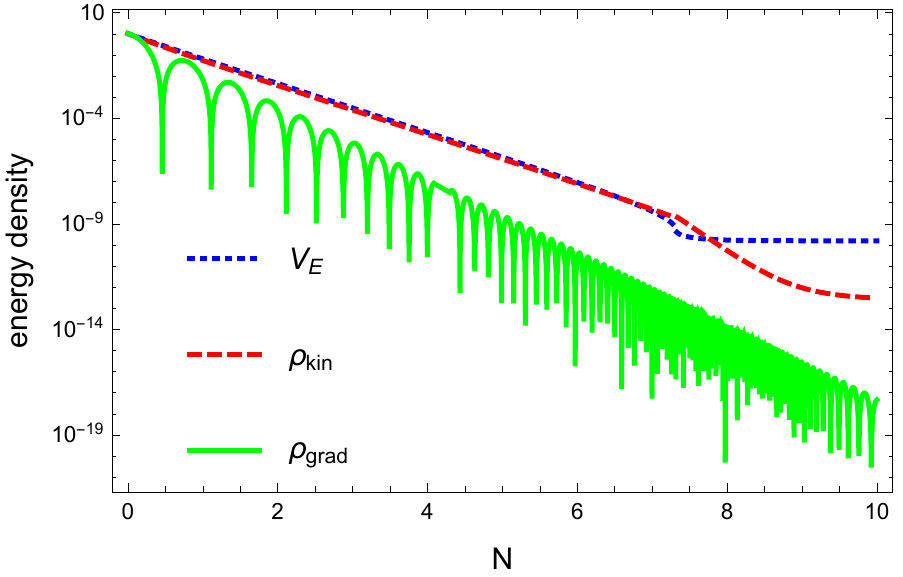}
\caption{\label{fig:rho}
Evolution of the gradient energy density $\rho_{{\rm grad}} \equiv (\nabla\varphi)^{2}/2$ (solid green), kinetic energy density $\rho_{{\rm kin}} \equiv \dot{\varphi}^{2}/2$ (dashed red) and potential energy density $V_{{\rm E}}$ (dotted blue).
After a short period of fast roll, which corresponds to $\epsilon_{H} \gtrsim 1$ (see Fig. \ref{fig:epsH}), the potential energy density dominates, allowing slow-roll inflation.
The same parameter values are chosen as in Fig. \ref{fig:epsH}.
}
\end{figure}

The kinetic energy density $\rho_{{\rm kin}}$, gradient energy density $\rho_{{\rm grad}}$ and the potential energy density are shown in Fig.~\ref{fig:rho}. One can see that the potential energy density starts to dominate after $N \approx 8$, at which point $\epsilon_{H}$ becomes less than unity when the field value $\overline{\varphi} \approx 9.76M_{{\rm P}}$. In general, the kinetic energy density is only slightly larger than the potential energy density during the fast roll phase. The gradient energy density, on the other hand, rapidly becomes small relative to the potential energy density\footnote{The oscillations of the gradient energy density in Fig.~\ref{fig:rho} are due to the transfer of the energy of the oscillating mode $\delta \varphi(\bfx, t)$ between gradient and kinetic energy.}.

\begin{figure}[t]
\centering
\includegraphics[width=0.45\textwidth]{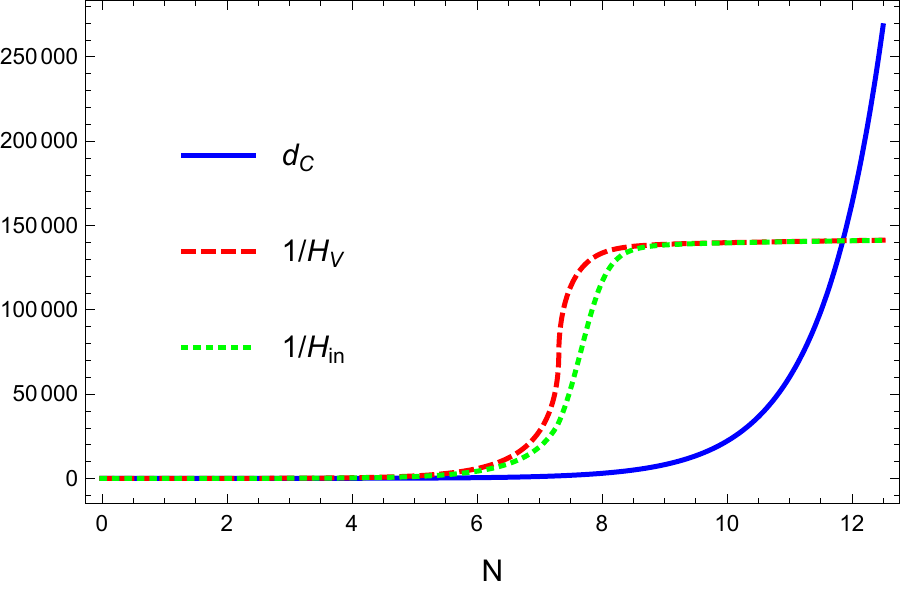}
\caption{\label{fig:horizon}
Three different horizon scales: the diameter of the classically evolving volume $d_{{\rm c}} \equiv a/M_{{\rm P}}$ (solid blue), the Hubble radius calculated using the potential energy  $H_{V}^{-1}$ (dashed red), and the Hubble radius of the classical volume $H_{{\rm in}}^{-1}$ (dotted green).
When $N \approx 12$, the classical volume becomes larger than the horizon scale, thus providing the smooth potential-dominated initial conditions for inflation.
The same parameter values are chosen as in Fig. \ref{fig:epsH}.
}
\end{figure}

During the initial noninflationary fast-rolling phase, the horizon may grow more rapidly than the diameter $d_{{\rm c}}$ of the classically evolving volume. Therefore we need to check that the horizon can become smaller than $d_{{\rm c}}$ after the onset of potential domination at $N \approx 8$. In Fig.~\ref{fig:horizon}, we show three different scales: $d_{{\rm c}} \equiv a M_{{\rm P}}^{-1}$, with $M_{{\rm P}}^{-1}$ being the natural horizon scale at the Planck initial state; the Hubble radius $H_{V}^{-1}$ calculated from the potential energy density [$H_{V} \equiv \sqrt{V_{{\rm E}}/(3M_{{\rm P}}^{2})}$]; and the Hubble radius $H_{{\rm in}}^{-1}$ calculated using the energy density inside the classically evolving volume.
As expected, $H_{V}^{-1} \approx H_{{\rm in}}^{-1}$ once the potential slow-roll is established after $N \approx 8$.
At $N \approx 12$, the classical volume becomes larger than the horizon scale, providing the smooth potential-dominated initial state required for inflation.

Thus we can conclude that when the scalar potential satisfies the slow-roll conditions, the potential energy will dominate the gradient energy density and kinetic energy density. Therefore there will be a smooth transition to potential-dominated slow-roll inflation. This will provide the initial conditions for the subsequent era of plateau inflation.

%---------------------
%---------------------

%---------------------
%---------------------

%---------------------
%---------------------

%%%%%%%%%%%%%%%%%%%%%%%%%%%%%%%%%%%%%%%%%%%%%%%%
\section{Cosmological observables in models with a conformal factor zero}
\label{sec:NMpoleCOs}
%%%%%%%%%%%%%%%%%%%%%%%%%%%%%%%%%%%%%%%%%%%%%%%%

\begin{figure*}[t]
\centering
\includegraphics[width=0.45\textwidth]{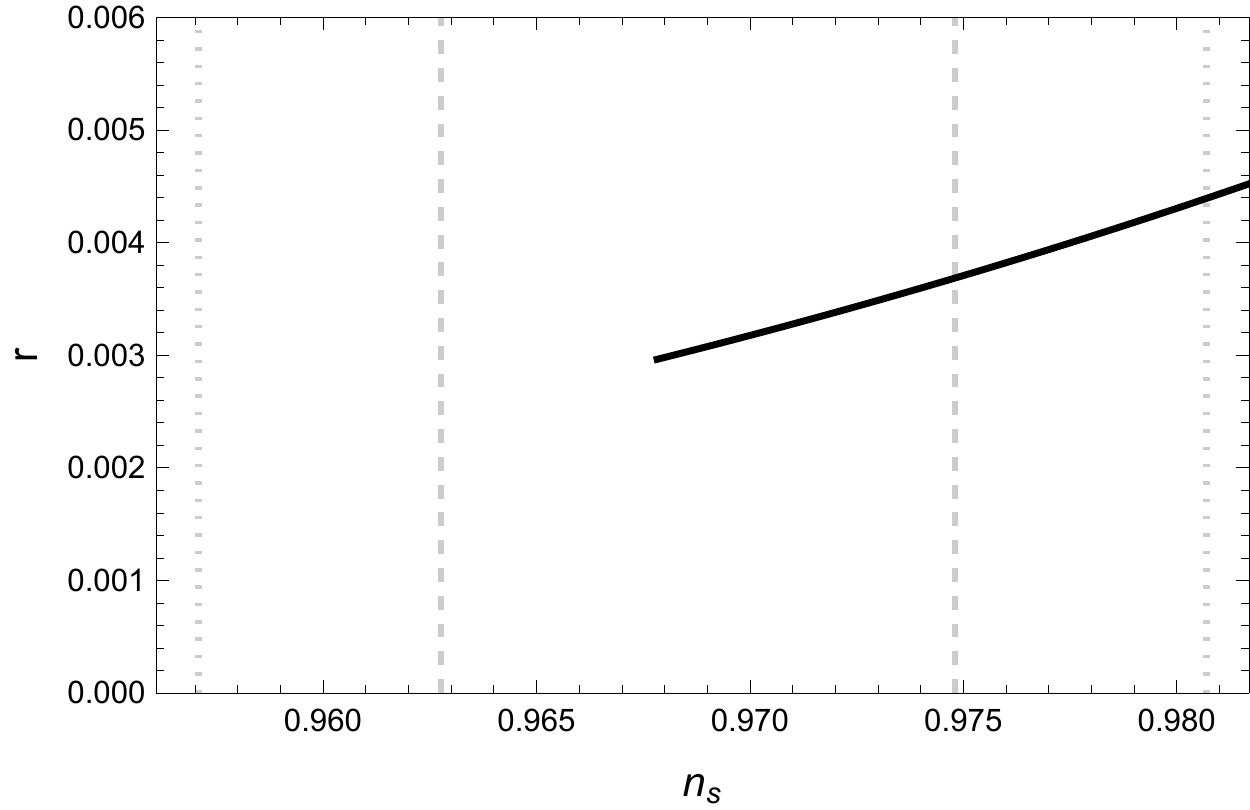}\;\;\;\;
\includegraphics[width=0.45\textwidth]{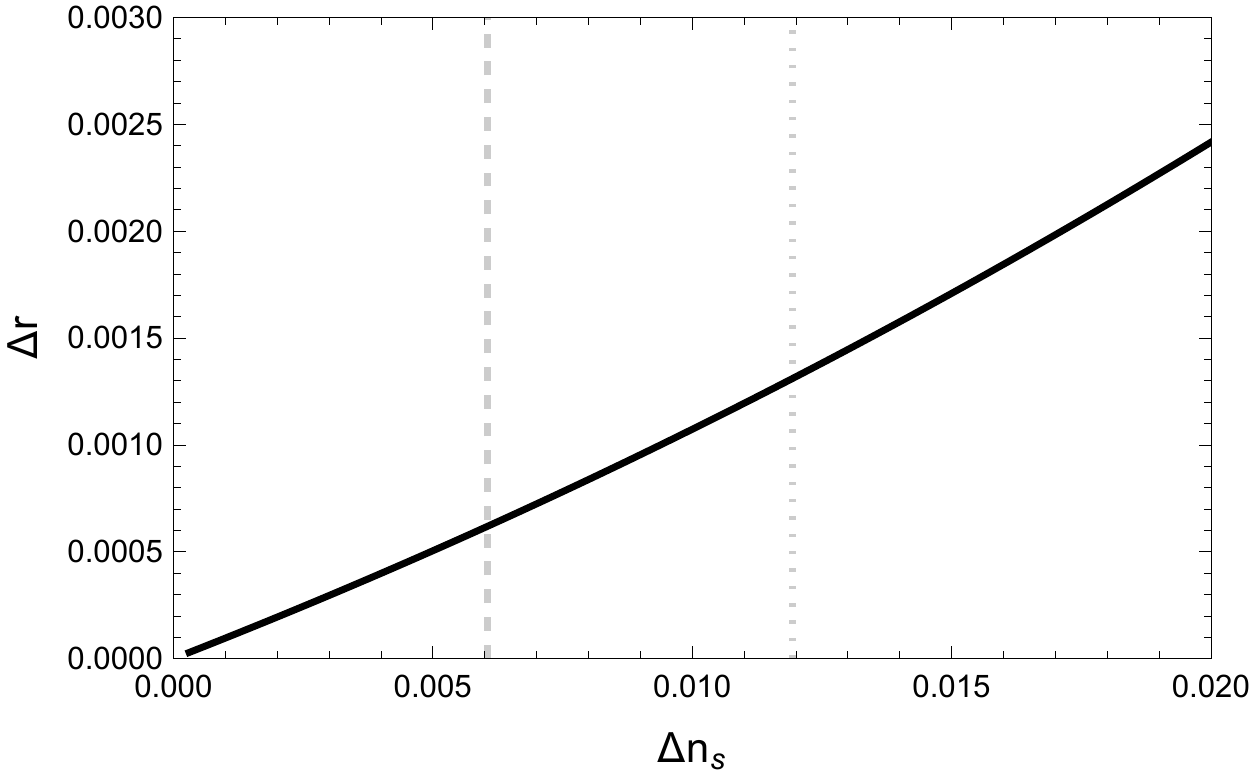}
\caption{\label{fig:nsr}
$n_{s}$ versus $r$ (left) and $\Delta n_{s}$ versus $\Delta r$ (right). 
The vertical dashed (dotted) lines corresponds to $1-\sigma$ ($2-\sigma$) Planck bound on the spectral index $n_{s}$.
The correlation between $n_{s}$ and $r$ is purely determined by the single free parameter, $\xi_{4}$, of the conformal zero models at small $\phi$.
In our analysis, the additional parameter $\xi_{4}$ is treated 
as a free parameter while $\xi_{2}$ becomes as function of $\xi_{4}$, being chosen in such a way that the Planck normalization on ${\cal P}_{s}$ is satisfied at $N =60$. The correlation is independent of $\lambda$.
}
\end{figure*}

The class of models described by Eq.~\eqref{eqn:generalcase} all reduce to the same conformal factor~\footnote{
An analysis of $n_{s}$ and $r$ for a general conformal factor expansion is presented in Ref.~\cite{Broy:2016rfg}.
} at small $\phi$,
\begin{align}
	\Omega^{2}
	\approx 
	1 + \xi_{2} \frac{\phi^{2}}{M_{{\rm P}}^{2}}
	-\xi_{4}\frac{\phi^{4}}{M_{{\rm P}}^{4}}
	\,,
\end{align}
where we expect $\xi_{2} \sim \xi_{4}$. Since the small $\phi$ limit of these models introduces only a single parameter $\xi_{4}$ (which must be greater than zero), this class of model predicts a specific correlation between the modification of $n_{s}$ and $r$. The first prediction is that both $n_{s}$ and $r$ strictly increase relative to the predictions of the standard model, due to the $\xi_{4}$ term being strictly positive. In Fig. \ref{fig:nsr} we show $n_{s}$ versus $r$ and $\Delta n_{s}$ versus $\Delta r$, where $\Delta n_{s} \equiv n_{s} - n_{s}^{{\rm ST}}$ and $\Delta r \equiv r - r^{{\rm ST}}$, with $n_{s}^{{\rm ST}}$ and $r^{{\rm ST}}$ being respectively the spectral index and tensor-to-scalar ratio of the standard nonminimal model \eqref{eqn:NMstandardCOs} (see also the Appendix).
For values of $n_{s}$ at the Planck $1-\sigma$ ($2-\sigma$) bound, the shift of $r$ from the standard nonminimal model value \eqref{eqn:NMstandardCOs} is by $\Delta r = 0.0006$ ($0.0013$). 
These shifts are larger than the projected accuracy of the next generation cosmic microwave background (CMB) polarization experiments \cite{Matsumura:2013aja, Kogut:2011xw, Remazeilles:2015hpa}, which are expected to ideally achieve an error $\delta r \approx 0.0002$ at $1-\sigma$. Therefore the observation of a small increase of $r$ above its standard nonminimal model value would be consistent with models with a conformal factor zero. Moreover, if the spectral index can be determined with increased precision then it may be possible to test the specific correlation between the shifts of $n_{s}$ and $r$ predicted by this class of models.

%---------------------
%---------------------

\begin{figure*}[t]
\centering
\includegraphics[width=0.45\textwidth]{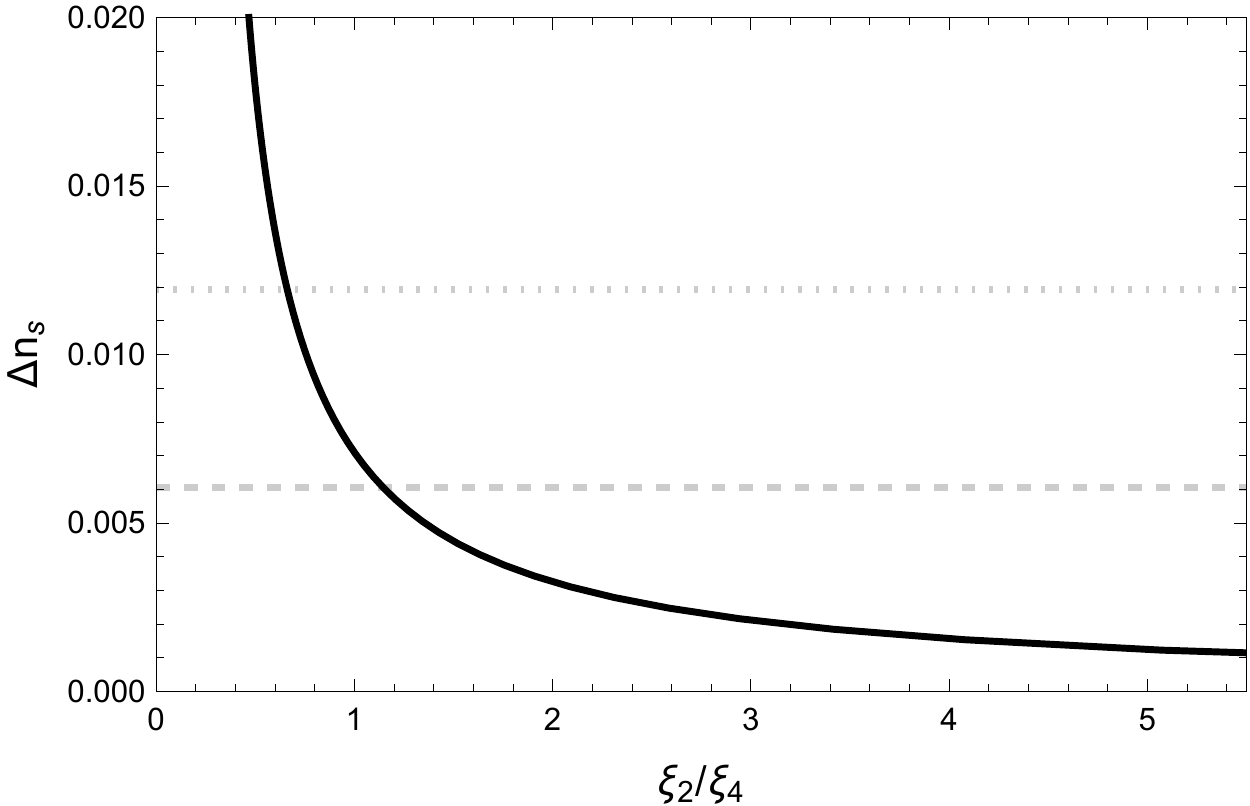}\;\;\;\;\;
\includegraphics[width=0.45\textwidth]{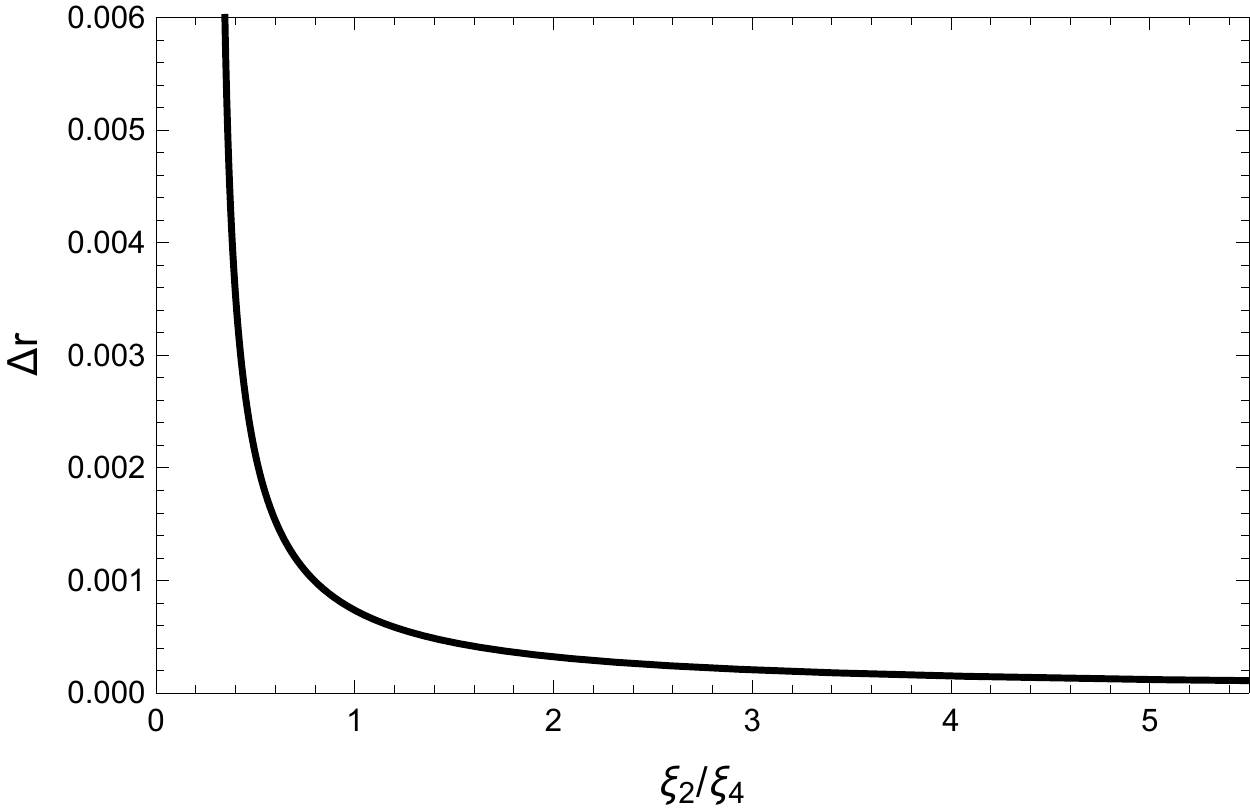}
\caption{\label{fig:nsxirxi}
$\Delta n_{s}$ (left) and $\Delta r$ (right) in terms of $\xi_{2}/\xi_{4}$. The horizontal dashed (dotted) line corresponds to $1-\sigma$ ($2-\sigma$) Planck bound on the spectral index $n_{s}$.
The $\xi_{4}=0$ limit corresponds to the standard nonminimal model \eqref{eqn:NMstandardCOs}.
An observable shift of $r$ in next generation CMB polarization experiments \cite{Matsumura:2013aja, Kogut:2011xw, Remazeilles:2015hpa}, $\Delta r \gtrsim 0.0002$, is possible when $\xi_{2}/\xi_{4}\lesssim 3$.
As in Fig. \ref{fig:nsr}, the additional parameter $\xi_{4}$ is treated as a free parameter and $\xi_{2}$ is adjusted so that the Planck normalization of ${\cal P}_{s}$ is satisfied at $N = 60$, in which case $\xi_{2}/\xi_{4}$ becomes a function of $\xi_{4}$.  $\lambda = 1$ is assumed in our analysis.}
\end{figure*}

In order to have an observable shift of $r$, $\xi_4$ must be within a particular range of values. For $\lambda = 1$, this corresponds to $0.6 \lesssim \xi_{2}/\xi_{4} \lesssim 3$, as seen in Fig. \ref{fig:nsxirxi}. Lower values of $\xi_{2}/\xi_{4}$ would produce an excessive increase in $n_{s}$ beyond the present Planck $2-\sigma$ upper bound, while larger values would produce unobservably small shifts of $r$.

\begin{figure*}[t]
\centering
\includegraphics[width=0.45\textwidth]{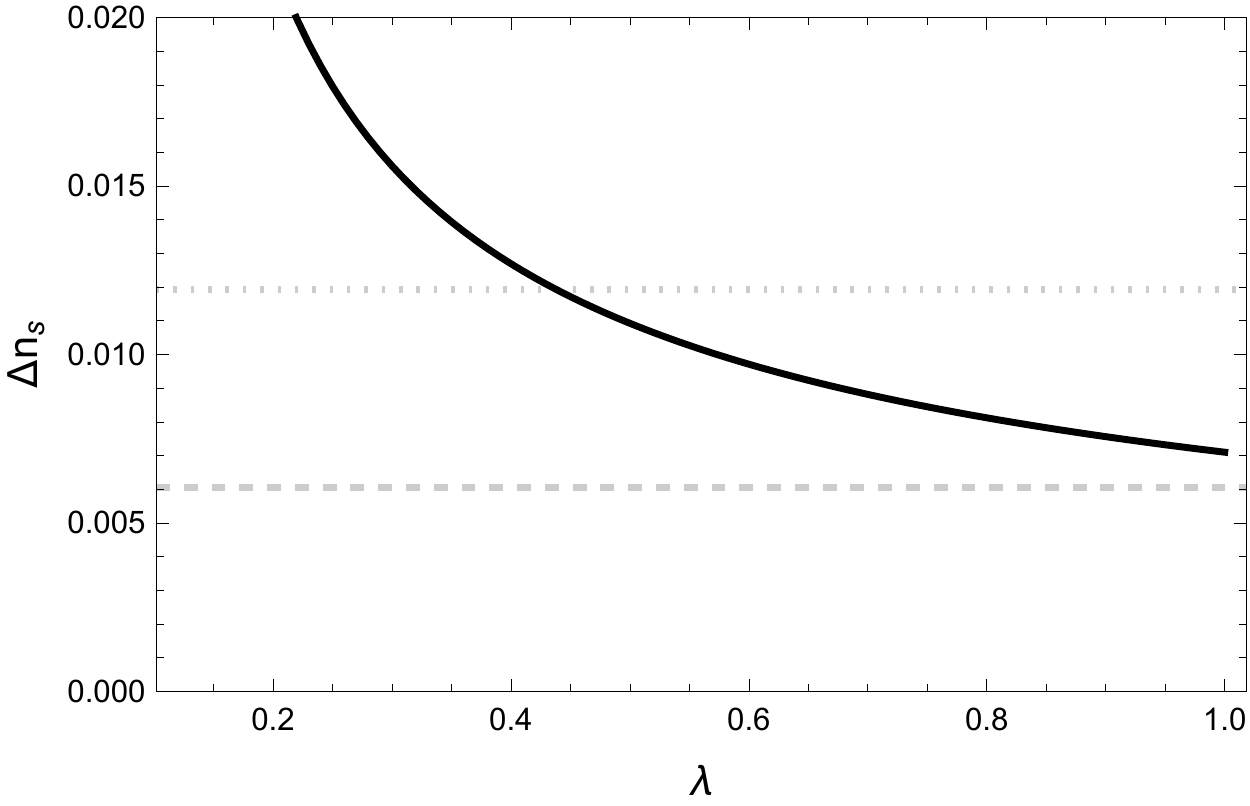}
\;\;\;\;
\includegraphics[width=0.45\textwidth]{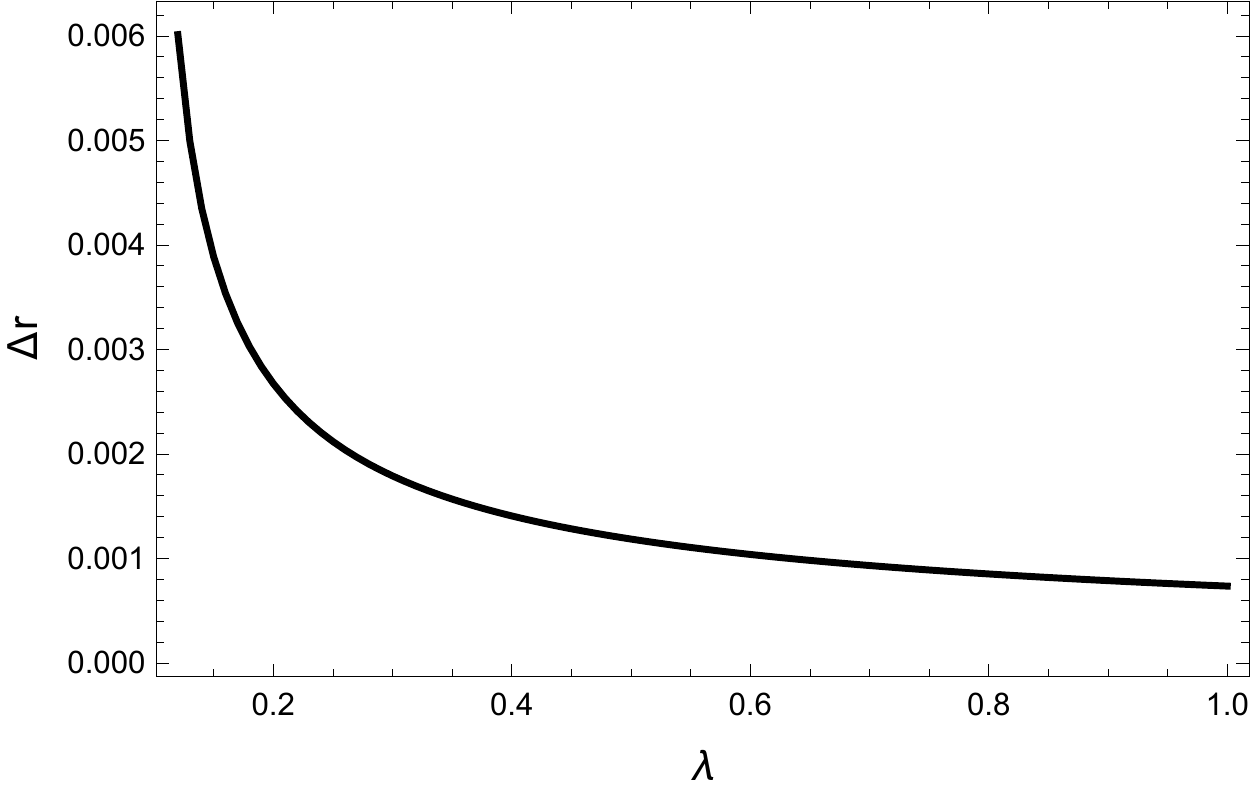}
\caption{\label{fig:xi2exi4}
$\Delta n_{s}$ and $\Delta r$ as a function of $\lambda$ in the case of $\xi_{2} = \xi_{4}$.
The horizontal dashed (dotted) line corresponds to $1-\sigma$ ($2-\sigma$) Planck bound on the spectral index $n_{s}$. Smaller values of $\lambda$ produce larger shifts of $n_{s}$ and $r$, and $\lambda \gtrsim 0.43$ is necessary in order that $n_{s}$ is within the present Planck $2-\sigma$ bound. We treat $\lambda$ as a free parameter while $\xi_{2}$ is chosen in such a way that the Planck normalization of ${\cal P}_{s}$ is satisfied at $N = 60$.
}
\end{figure*}

The dependence of $\Delta n_{s}$ and $\Delta r$ on $\lambda$ are illustrated in Fig. \ref{fig:xi2exi4} for the case $\xi_{4} = \xi_{2}$. $\lambda \gtrsim 0.43$ is necessary in order that $\Delta n_{s}$ is within the present Planck $2-\sigma$ bound; smaller values of $\lambda$ produce larger shifts of $n_s$ and $r$.

%---------------------
%---------------------

%---------------------
%---------------------

%%%%%%%%%%%%%%%%%%%%%%%%%%%%%%%%%%%%%%%%%%%%%%%%
\section{Conclusions}
\label{sec:con}
%%%%%%%%%%%%%%%%%%%%%%%%%%%%%%%%%%%%%%%%%%%%%%%%
Nonminimally coupled scalar inflation is in excellent agreement with observation, but it requires an explanation of how inflation got started in the first place. This is most easily understood if the Universe emerges from a chaotic initial state with Planck-scale energy density. However, this is not possible for the standard nonminimally coupled inflation model, as it is a plateau inflation model with $V_{{\rm E}} \ll M_{{\rm P}}^4$ when expressed in the Einstein frame. 

By modifying the conformal factor of the standard model to a conformal factor with a zero, it is possible to achieve a Planck potential energy density.  We have proposed a class of models which does this by multiplying the nonminimal coupling term $\xi_{2} \phi^2/M_{{\rm P}}^2$ by a factor $f(\phi^2/M_{{\rm P}}^2)$ which tends to 1 at small $\phi$, so preserving the successful predictions of the standard model, while making the conformal factor equal zero at large $\phi$. The use of a conformal factor with a zero may be considered to be a minimal modification of the original model, in the sense that it modifies only its coupling to gravity and does not modify the particle physics model itself.  

In the case of the simplest example of a model with a conformal factor with a zero and chaotic initial conditions, we showed that the model can smoothly evolve into slow-roll inflation, which later evolves into plateau inflation. There is a brief period, $\Delta N \approx 8$,  of fast-roll noninflationary expansion following the initial chaotic era. However, the gradient and kinetic energy densities never strongly dominate the potential energy density, and the potential energy comes to dominate by the time the scalar potential satisfies the slow-roll conditions. 

In general, the class of models we are considering predicts a correlation between the deviation of $n_{s}$ and $r$ from their standard nonminimal model values which is independent of the specific form of $f(\phi^{2}/M_{{\rm P}}^{2})$. In particular, the model predicts that $n_s$ and $r$ can only increase relative to their standard values. If the single relevant additional parameter of the models at small $\phi$, $\xi_{4}$, is of the right magnitude, then an increase of $r$ by as much as 0.0013 is possible when $n_{s}$ is within the $2-\sigma$ upper bound observed by Planck. It turns out that $\xi_{4}$ can produce shifts of $r$ which are large enough to be observed by future CMB satellites if $\xi_{4} \sim \xi_{2}$ and $\lambda \sim 1$, where $\xi_{2} \sim 10^{4}$ is the nonminimal coupling of the standard nonminimally coupled inflation model and $\lambda$ is the $\phi^4$ coupling constant. Remarkably, $\xi_{4} \sim \xi_{2}$ is the natural expectation in the class of models we have proposed. Therefore an observable increase in $r$, correlated with an increase in $n_{s}$, is a natural possibility in these models.

%%%%%%%%%%%%%%%%%%%%%%%%%%%%%%%%%%%%%%%%%%%%%%%%
%\section*{Acknowledgements}
%%%%%%%%%%%%%%%%%%%%%%%%%%%%%%%%%%%%%%%%%%%%%%%%
%The work of JM was partly supported by STFC via the 
%Lancaster-Manchester-Sheffield Consortium for Fundamental 
%Physics under STFC grant ST/J000418/1. 

%%%%%%%%%%%%%%%%%%%%%%%%%%%%%%%%%%%%%%%%%%%%%%%%
%%%%%%%%%%%%%%%%%%%%%%%%%%%%%%%%%%%%%%%%%%%%%%%%
\section*{Appendix: ANALYTIC EXPRESSIONS FOR THE COSMOLOGICAL OBSERVABLES} 
\label{apdx:NMPapprox}
\renewcommand{\theequation}{A-\arabic{equation}}
 % redefine the command that creates the equation no.
 \setcounter{equation}{0}
% reset counter
%%%%%%%%%%%%%%%%%%%%%%%%%%%%%%%%%%%%%%%%%%%%%%%
%%%%%%%%%%%%%%%%%%%%%%%%%%%%%%%%%%%%%%%%%%%%%%%
In this appendix we obtain approximate analytic expressions for the cosmological observables in our model. Let us first write the conformal factor as follows:
\begin{align}
	\Omega^{2}
	=
	\xi_{2}\frac{\phi^{2}}{M_{{\rm p}}^{2}}(1+\delta)
	\,,\qquad
	\delta = \frac{M_{{\rm P}}^{2}}{\xi_{2}\phi^{2}}
	-\frac{\xi_{4}\phi^{2}}{\xi_{2}M_{{\rm P}}^{2}}
	\,,
\end{align}
where $\delta$ shall be treated as a small perturbation during inflation, \textit{i.e.}, $|\delta|\ll 1$. Note that 
\begin{align}
	\mathcal{O}(\delta) \sim
	\mathcal{O}(\phi \delta_{\phi}) \sim
	\mathcal{O}(\phi^{2} \delta_{\phi\phi})\,,
\end{align}
where $\delta_{\phi}\equiv d\delta/d\phi$ and so on. In terms of $\delta$, the Einstein-frame potential $V_{{\rm E}}$ and the relation between the canonically normalized field $\varphi$ and the original field $\phi$ are given by
\begin{align}
	V_{{\rm E}} &\approx
	\frac{\lambda M_{{\rm P}}^{4}}{4\xi_{2}^{2}}\left[
		1 - 2\delta + 3\delta^{2}
	\right]
	\,,\\
	\left(
		\frac{d\varphi}{d\phi}
	\right)^{2}
	&\approx
	\frac{6M_{{\rm P}}^{2}}{\phi^{2}}\left[
		1 + \phi\delta_{\phi}
		+\frac{1}{4}(\phi\delta_{\phi})^{2}
		-\delta\phi\delta_{\phi}
	\right]\,,
\end{align}
where we have used $\xi_{2} \gg 1$.

The slow-roll parameters are given by
\begin{align}
	\epsilon &\approx
	\frac{1}{3}(\phi\delta_{\phi})^{2}\left(
		1 - 2\delta - \phi\delta_{\phi}
	\right)
	\nonumber\\
	&=
	\frac{4}{3\xi_{2}^{2}}
	\left(
		\frac{M_{{\rm P}}}{\phi}
	\right)^{4}
	\left(
		1 + \xi_{4}\frac{\phi^{4}}{M_{{\rm P}}^{4}}
	\right)^{2}\left(
		1 + \frac{4\xi_{4}}{\xi_{2}}\frac{\phi^{2}}{M_{{\rm P}}^{2}}
	\right)
	\,,\label{eqn:epsilonPot}
\end{align}
\begin{align}
	\eta &\approx
	\frac{1}{3}\bigg[
		-\phi\delta_{\phi}
		-\phi^{2}\delta_{\phi\phi}
		\nonumber\\
		&\qquad
		+\frac{9}{2}(\phi\delta_{\phi})^{2}
		+\delta\phi^{2}\delta_{\phi\phi}
		+\delta\phi\delta_{\phi}
		+\frac{3}{2}\phi\delta_{\phi}\phi^{2}\delta_{\phi\phi}
	\bigg]
	\nonumber\\
	&=
	-\frac{4}{3\xi_{2}}\left(
		\frac{M_{{\rm P}}}{\phi}
	\right)^{2}\bigg[
		1 - \xi_{4}\frac{\phi^{4}}{M_{{\rm P}}^{4}}
		\nonumber\\
		&\qquad
		-\frac{1}{\xi_{2}}\left(
			\frac{M_{{\rm P}}^{2}}{\phi^{2}}
			+4\xi_{4}\frac{\phi^{2}}{M_{{\rm P}}^{2}}
			+7\xi_{4}^{2}\frac{\phi^{6}}{M_{{\rm P}}^{6}}
		\right)
	\bigg]
	\,.
\end{align}
Assuming the natural value for $\xi_{4}$, $\xi_{4} \sim \xi_{2}$, the number of $e$-folds $N$ is given by
\begin{align}
	N \approx
	\frac{3\xi_{2}}{4\sqrt{\xi_{4}}}\left[
		\arctan\left(
			\sqrt{\xi_{4}}
			\frac{\phi_{N}^{2}}{M_{{\rm P}}^{2}}
		\right)
		-
		\arctan\left(
			\sqrt{\xi_{4}}
			\frac{\phi_{{\rm e}}^{2}}{M_{{\rm P}}^{2}}
		\right)
	\right]\,,
\end{align}
where $\phi_{{\rm e}}$ is the field value at the end of inflation set by $\epsilon \simeq 1$ which is given by
\begin{align}
	\phi_{{\rm e}}
	&\approx
	\frac{M_{{\rm P}}}{2^{1/4}\sqrt{\xi_{4}}}
	\left[
		\frac{3\xi_{2}^{2}}{4} - 2\xi_{4}
		- \sqrt{
		\frac{3\xi_{2}^{2}}{4}\left(
			\frac{3\xi_{2}^{2}}{4} - 4\xi_{4}
		\right)
		}
	\right]^{1/4}
	\nonumber\\
	&\approx
	\left(
		\frac{4}{3}
	\right)^{1/4}\frac{M_{{\rm P}}}{\sqrt{\xi_{2}}}
	\,.
\end{align}
It is then easy to show that
\begin{align}
	\phi_{N} &\approx
	\frac{M_{{\rm P}}}{\xi_{4}^{1/4}}\left(
		\tan\left[
			\frac{4\sqrt{\xi_{4}}}{3\xi_{2}}N
			+\arctan\left(
				\sqrt{\xi_{4}}\phi_{{\rm e}}^{2}/M_{{\rm P}}^{2}
			\right)
		\right]
	\right)^{1/2}
	\nonumber\\
	&\approx
	\left(
		\frac{4N}{3}
	\right)^{1/2}
	\frac{M_{{\rm P}}}{\sqrt{\xi_{2}}}
	\,.
	\label{eqn:apdxphiN}
\end{align}
Note that the approximated expressions for $\phi_{{\rm e}}$ and $\phi_{N}$ are the same as those in the standard nonminimal inflation model.

The cosmological observables are then given by
\begin{align}
	\mathcal{P}_{s} &\approx
	\frac{\lambda}{128\pi^{2}\xi_{2}}\left(
		\frac{\phi}{M_{{\rm P}}}
	\right)^{2}\left[
		\frac{-2+\xi_{2}\phi^{2}/M_{{\rm P}}^{2}-2\xi_{4}\phi^{4}/M_{{\rm P}}^{4}}{(1 + \xi_{4}\phi^{4}/M_{{\rm P}}^{4})^{2}}
	\right]
	\,,\\
	n_{s} &\approx
	\frac{1}{3\xi_{2}^{2}}\left(
		\frac{M_{{\rm P}}}{\phi}
	\right)^{4}\bigg[
		-16-8\xi_{2}\frac{\phi^{2}}{M_{{\rm P}}^{2}}
		\nonumber\\
		&\qquad
		+(3\xi_{2}^{2}-16\xi_{4})\frac{\phi^{4}}{M_{{\rm P}}^{4}}
		+8\xi_{2}\xi_{4}\frac{\phi^{6}}{M_{{\rm P}}^{6}}
		+32\xi_{4}^{2}\frac{\phi^{8}}{M_{{\rm P}}^{8}}
	\bigg]
	\,,\\
	r &\approx
	\frac{64}{3\xi_{2}^{2}}\left(
		\frac{M_{{\rm P}}}{\phi}
	\right)^{4}\left(
		1 + \xi_{4}\frac{\phi^{4}}{M_{{\rm P}}^{4}}
	\right)^{2}\left(
		1 + \frac{4\xi_{4}}{\xi_{2}}\frac{\phi^{2}}{M_{{\rm P}}^{2}}
	\right)
	\,,
\end{align}
where the above expressions are evaluated at $\phi = \phi_{N}$.
Substituting $\phi_{N}$ \eqref{eqn:apdxphiN} into the above expressions for the cosmological observables gives
\begin{align}
	\mathcal{P}_{s}
	&\approx
	0.0014 N^{2} \times \frac{\lambda}{\xi_{2}^{2}}
	\frac{1-2.667 N \xi_{4}/\xi_{2}^{2}}{\left(
		1+1.778 N^{2} \xi_{4}/\xi_{2}^{2}
	\right)^{2}}
	\nonumber\\
	&\approx
	\mathcal{P}_{s}^{{\rm ST}} \times
	\frac{1-2.667 N \xi_{4}/\xi_{2}^{2}}{\left(
		1+1.778 N^{2} \xi_{4}/\xi_{2}^{2}
	\right)^{2}}
	\,,
\end{align}
\begin{align}
	n_{s}
	&\approx
	1 - \frac{2}{N} - \frac{3}{N^{2}}
	+\frac{3.556 \, N \xi_{4}}{\xi_{2}^{2}}\left(
		1 + 5.33 N \frac{\xi_{4}}{\xi_{2}^{2}}
	\right)
	\nonumber\\
	&\approx
	n_{s}^{{\rm ST}}
	+ \frac{3.556 \, N \xi_{4}}{\xi_{2}^{2}}\left(
		1 + 5.33 N \frac{\xi_{4}}{\xi_{2}^{2}}
	\right)
	\,,
      \end{align}
     \begin{align}   
     r
	&\approx
	\frac{12}{N^{2}}
	+\times \frac{42.67 \,\xi_{4}}{\xi_{2}^{2}}\left(
		1 + 0.889 N^{2} \frac{\xi_{4}}{\xi_{2}^{2}}
		+4.74 N^{3} \frac{\xi_{4}^{2}}{\xi_{2}^{4}}
	\right)
	\nonumber\\
	&\approx
	r^{{\rm ST}}
	+\frac{42.67\, \xi_{4}}{\xi_{2}^{2}}\left(
		1 + 0.889 N^{2} \frac{\xi_{4}}{\xi_{2}^{2}}
		+4.74 N^{3} \frac{\xi_{4}^{2}}{\xi_{2}^{4}}
	\right)
	\,, 
\end{align}
where quantities with the superscript ST are those of the standard nonminimal model.
Therefore the deviations from the standard nonminimal case are given by
\begin{align}
	\Delta n_{s}
	&\approx
	 \frac{3.556\, N \xi_{4}}{\xi_{2}^{2}}\left(
		1 + 5.33 N \frac{\xi_{4}}{\xi_{2}^{2}}
	\right)
	\,,
\end{align}
\begin{align} 
	\Delta r
	&\approx
	\frac{42.67 \, \xi_{4}}{\xi_{2}^{2}}\left(
		1 + 0.889 N^{2} \frac{\xi_{4}}{\xi_{2}^{2}}
		+4.74 N^{3} \frac{\xi_{4}^{2}}{\xi_{2}^{4}}
	\right)
	\,.
\end{align}
We find that these expressions are in good agreement with the exact numerical values. It is then easy to see why $\xi_{4} \approx \xi_{2} \approx 5 \times 10^{4} \sqrt{\lambda}$ produces a shift $\Delta r \sim 0.001$ when $\lambda \sim 1$ and $N = 60$.

%%%%%%%%%%%%%%%%%%%%%%%%%%%%%%%%%%%%%%%%%%%%%%%
%%%%%%%%%%%%%%%%%%%%%%%%%%%%%%%%%%%%%%%%%%%%%%%

%%%%%%%%%%%%%%%%%%%%%%%%%%%%%%%%%%%%%%%%%%%%%%%
%%%%%%%%%%%%%%%%%%%%%%%%%%%%%%%%%%%%%%%%%%%%%%%

\end{document}